\documentclass{article}

\PassOptionsToPackage{numbers, compress}{natbib}

\usepackage[preprint]{style_templete}

\usepackage[utf8]{inputenc} 
\usepackage[T1]{fontenc}    
\usepackage{hyperref}       
\usepackage{url}            
\usepackage{booktabs}       
\usepackage{nicefrac}       
\usepackage{microtype}      
\usepackage{enumitem}
\usepackage{float}
\usepackage{algorithm}
\usepackage{xcolor}

\usepackage{algorithmic}

\usepackage{graphicx,subfigure}
\usepackage{amsmath,amsfonts}
\usepackage{multirow,soul}

\usepackage{placeins}
\usepackage{tabularx,threeparttable}
\usepackage{algorithm}
\usepackage{color}
\usepackage{epstopdf}
\usepackage{amsmath,amssymb}
\usepackage{setspace}

\usepackage{placeins}
\usepackage{tabularx,threeparttable}
\usepackage{algorithm}
\usepackage{color}
\usepackage{epstopdf}
\usepackage{amsmath,amssymb}
\usepackage{setspace}
\usepackage{mathrsfs}

\definecolor{maroon}{RGB}{122,000,25}
\definecolor{purple}{RGB}{122,000,250}

\title{SynCell: Contextualized Drug Synergy Prediction}

\small{

\author{
  Keqin Peng\thanks{University of Glasgow} \\
  \and
  Qinshan Shi \thanks{MBZUAI}\\
  \and
  Shuai Gao\thanks{Neurosurgery, Brigham and Women's Hospital} \\
  \and
  Guangxin Su\thanks{University of New South Wales} \\
  \and
  Ren Wang\thanks{Illinois Institute of Technology} \\
  \and
  Can Chen\thanks{University of North Carolina at Chapel Hill} \\
  \and
  Jun Wen\thanks{MBZUAI, Harvard Medical School}
}

}

\begin{document}

\maketitle

\begin{abstract}
\textbf{Motivation:}
Drug synergy is profoundly influenced by cellular context, as variations in protein interaction landscapes and pathway activities across cell types reshape how drugs act in combination. Most existing models overlook this heterogeneity, relying on static or bulk-level protein--protein interaction (PPI) networks that ignore cell-specific molecular wiring. The availability of large-scale transcriptomic data now enables the reconstruction of cell-line-resolved interactomes, offering a new foundation for contextualized drug synergy modeling.

\textbf{Results:}
Here we present SynCell, a Contextualized Drug Synergy framework that integrates drug--protein, protein--protein, and protein--cell line relations within a unified graph architecture. SynCell leverages cell-line-specific PPI networks to embed the molecular context in which drugs act, and employs graph convolutional learning to model how pharmacological effects propagate through cell-specific signaling networks. This formulation treats synergy prediction as a cell-line-contextualized drug--drug interaction problem. Across the large-scale DrugCombDB benchmark, SynCell consistently outperforms state-of-the-art baselines—including DeepSynergy, HypergraphSynergy, HERMES, BAITSAO, DTF, and NHP—particularly in predicting synergies involving unseen drugs or novel cell lines. When benchmarked against these seven methods, SynCell demonstrates substantial gains in generalization and biological interpretability, confirming that contextualizing PPIs with cell-line resolution is indispensable for accurate synergy prediction.
\end{abstract}

\section{Introduction}

Drug combination therapies represent a cornerstone of modern oncology, offering enhanced efficacy, reduced toxicity, and the ability to overcome resistance mechanisms~\cite{jia2009mechanisms, humphrey2011challenges}. Despite their clinical promise, the systematic identification of synergistic drug pairs is hindered by the vast combinatorial space, estimated to exceed millions of potential pairs, and pronounced heterogeneity in cellular responses~\cite{bray2024cancer, zhao2020combinatorial}. The urgency is further compounded by the prevalence of drug resistance, where tumor evolution renders monotherapies ineffective, necessitating multi-target interventions~\cite{holohan2013resistance}. Computational approaches have evolved significantly to address this challenge. Early efforts relied on feature-engineered machine learning models, such as random forests and matrix factorization, which integrated chemical descriptors and gene expression profiles to classify synergistic pairs~\cite{li2018randomforest, nafshi2021pmf}. While interpretable, these methods often suffer from limited capacity to capture high-order nonlinear interactions. The advent of deep learning enabled end-to-end learning from raw molecular inputs, pioneered by DeepSynergy~\cite{preuer2018deepsynergy}, which fused drug chemical structures with cell line transcriptomics using multi-layer perceptrons. Subsequent works introduced convolutional and recurrent architectures to better encode SMILES strings and gene sequences~\cite{wang2024deeplearningreview}, achieving notable success in predicting synergy scores, yet often relying on bulk-level data that masks cellular context~\cite{liu2020drugcombdb}.

To capture structural dependencies beyond flat feature vectors, recent approaches have adopted graph neural networks (GNNs) and knowledge graphs (KGs)~\cite{besharatifard2024gnnreview}. DeepDDS~\cite{wang2021deepdds} applied graph attention networks to encode drug molecular graphs alongside gene expression vectors, while GraphSynergy~\cite{yang2021graphsynergy} incorporated protein-protein interaction (PPI) networks via graph convolution to capture topological dependencies. Knowledge graph-based methods like KGANSynergy~\cite{zhang2023kgansynergy} further constructed unified graphs of drug-target-cell line relations to enhance semantic richness. Despite these progresses, a critical limitation persists: most existing methods employ static, pan-cancer PPI networks that ignore cell-type-specific signaling contexts. Drugs exert their effects by perturbing intracellular signaling pathways whose topologies are defined by context-aware PPI networks~\cite{menche2015uncovering, szklarczyk2023string}. Consequently, models that treat PPIs as static backbones or uniformly aggregate neighborhood information fail to capture the cell-specific molecular environments that critically influence drug synergy~\cite{cheng2024hansynergy, hao2025gnnsynergy}. This disconnect between static network priors and dynamic cellular states limits generalization to unseen cell lines and novel drug combinations, mirroring the zero-shot challenges observed in broader drug repurposing tasks~\cite{huang2024txgnn}.

Recognizing the importance of cellular context, recent works have explored context-aware architectures. CCSynergy~\cite{hosseini2023ccsynergy} introduced conditional neural modules to adapt drug representations based on cell line features, and other studies employed attention mechanisms to weight pathway activities according to tissue-specific expression~\cite{edwards2023contextlearning}. However, these methods typically modulate features at the representation level without explicitly modeling the underlying context-specific biological networks. As a result, they may miss mechanistic insights encoded in cell-type-specific PPI subnetworks. \textbf{Notably, expression-guided reweighting has emerged as a validated paradigm for context-specific interactome reconstruction, where transcriptomic profiles dynamically filter or weight edges in universal PPI networks \cite{greene2015understanding}. Recent updates to resources like STRING have further strengthened this approach by upgrading co-expression channels to support multi-omics data input for generating high-confidence context-specific subnetworks \cite{szklarczyk2023string}.} The maturation of large-scale biological databases now enables the construction of truly context-aware models. Publicly available resources on drug-target interactions~\cite{chandak2022primekg}, global PPIs~\cite{szklarczyk2023string}, and cell line expression profiles~\cite{arafeh2025depmap, liu2020drugcombdb} provide the foundation for building unified knowledge graphs that capture cell-specific interactions.

Here we introduce SynCell, a heterogeneous graph neural network framework for contextualized drug synergy prediction. Our work makes two fundamental advances over existing methods. First, we systematically incorporate protein-protein interaction networks not merely as interpretability tools but as core biological priors that encode functional relationships between drug targets. Second, and more importantly, we introduce contextualized drug-drug relation modelling through cell-specific PPI subnetworks and adaptive feature modulation.

Our core innovation involves three key components: (1) a unified knowledge graph embedding drugs, proteins, and cell lines through three biological relationships—drug-protein binding, protein-protein interactions, and protein-cell line associations; (2) dynamic construction of cell line-specific PPI subnetworks by integrating global PPIs with cell-specific expression profiles; and (3) an adaptive feature modulation mechanism that transforms drug representations based on cellular context through learnable scaling and shifting operations, ensuring distinct embeddings for the same drug across different environments and enabling contextualized drug-drug relation modelling.

Extensive evaluations demonstrate that SynCell achieves state-of-the-art performance across multiple benchmarks, with particularly pronounced advantages in clinically challenging scenarios involving novel drug combinations and unseen cell lines. Beyond providing a high-performance predictive tool, our work offers the first computational evidence that cell-specific PPI networks are indispensable for accurate synergy prediction. This insight establishes a new paradigm for precision oncology, enabling more accurate and interpretable design of personalized combination therapies.

\begin{figure}[htbp]
    \centering
    \includegraphics[width=1\linewidth]{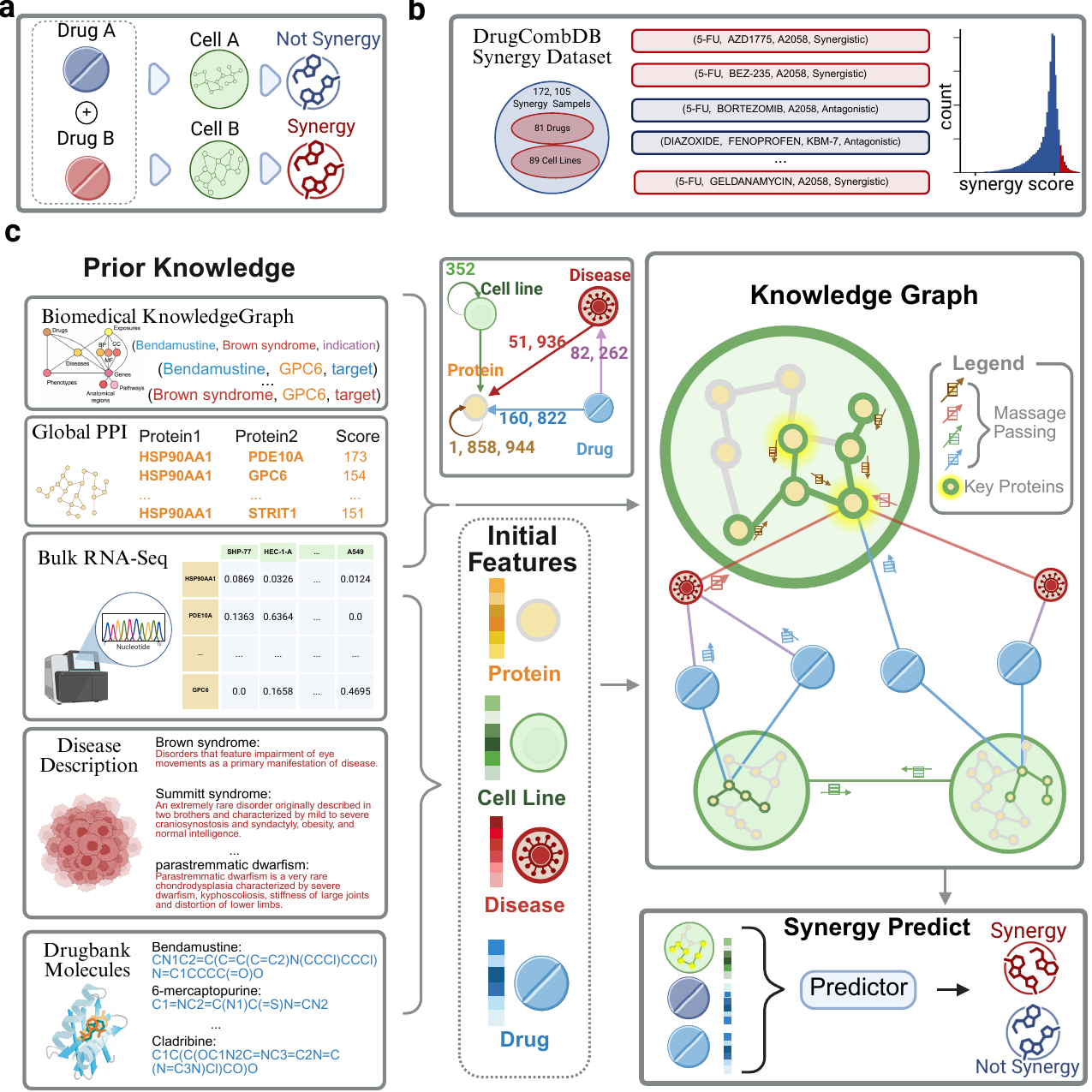}
\caption{
\textbf{SynCell is a contextualized graph framework for drug synergy prediction, leveraging cell-line-specific PPI networks to capture molecular heterogeneity.} 
\textbf{a,} Cellular context shapes drug response. Drug synergy varies across cell lines due to differences in protein activation contexts, where the same drug combination may exhibit synergistic effects in one cell line but not in another. 
\textbf{b,} Multimodal knowledge graph construction. SynCell integrates data from DrugCombDB, DepMap, STRING, and PrimeKG to build a unified heterogeneous graph. DrugCombDB provides synergy measurements; DepMap summarizes cell-line-specific protein expression profiles; STRING presents global protein–protein interactions; and PrimeKG illustrates biological relationships including drug–target and disease–protein associations. 
\textbf{c,} Computational framework overview. SynCell operates through three phases: (1) context-aware graph construction derives cell-line-specific PPI subnetworks from the global PPI based on activated genes; (2) multimodal embeddings are generated via SVD for expression data and transformer encoders (ChemBERTa, PubMedBERT) for drug and disease features; (3) a heterogeneous graph neural network integrates hyperedge propagation and adaptive contextual modulation to predict synergy. The model jointly optimizes synergy prediction and organ classification to enhance generalization across unseen biological environments.
}
    \label{fig:framework}
\end{figure} 
\section{Results}

\begin{figure}[htbp]
    \centering
    \includegraphics[width=1\linewidth]{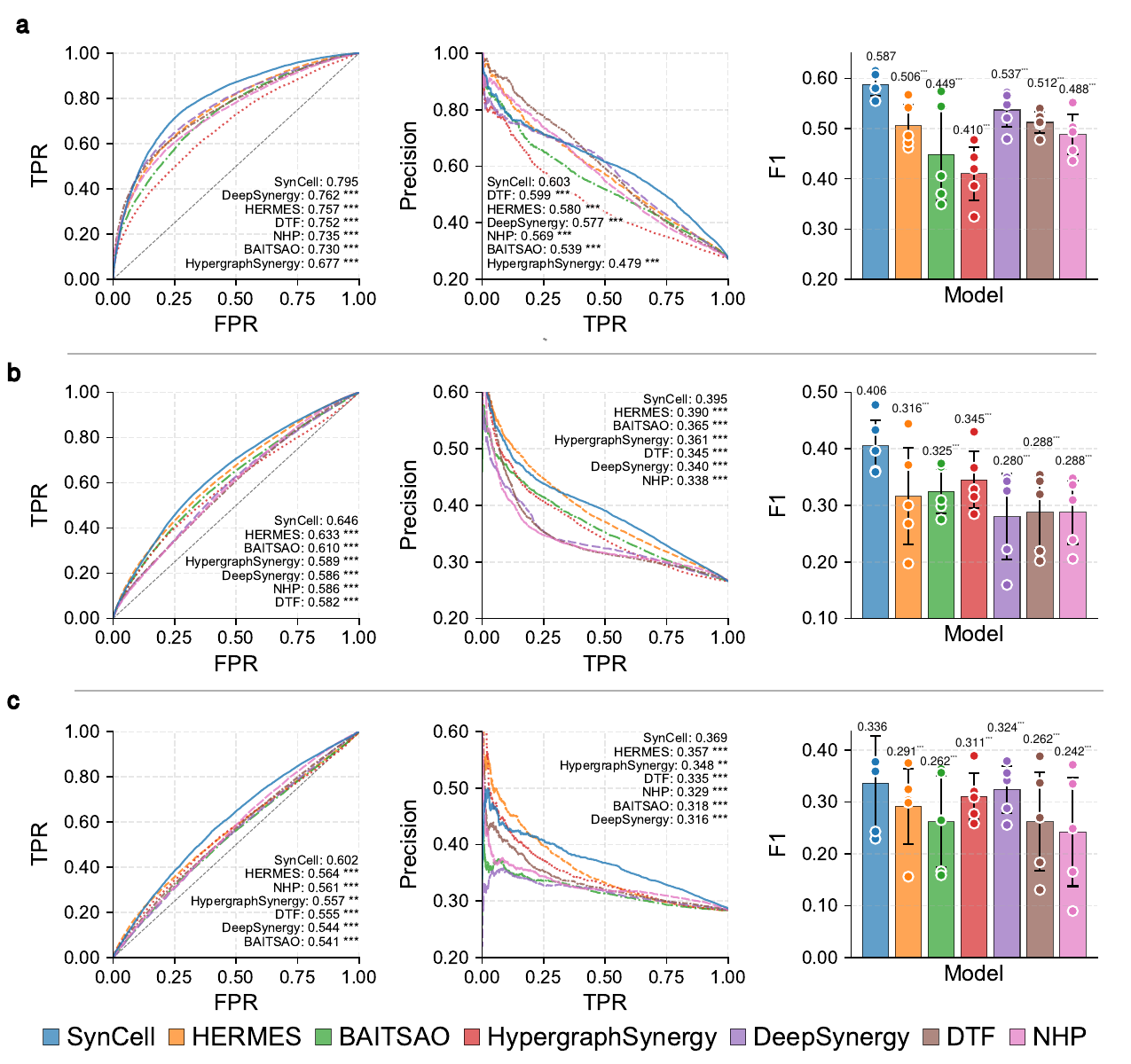}
    
    \caption{
    \textbf{SynCell achieves robust zero-shot generalization to unseen drugs and drug combinations across progressively challenging evaluation settings.} 
    \textbf{a,} DrugComb split: generalization to unseen drug combinations while all individual drugs are observed during training. SynCell achieves the highest AUROC (0.795), AUPRC (0.603) and F1 (0.587), outperforming DeepSynergy, HERMES and other baselines. 
    \textbf{b,} DrugSingle split: semi-cold-start scenario where one drug in each test pair is entirely unseen. SynCell maintains superior ranking consistency (AUROC: 0.646) and classification balance (F1: 0.406) despite increased task difficulty (AUPRC: 0.395). 
    \textbf{c,} DrugDouble split: fully cold-start scenario where both drugs in test pairs are absent from training. Although performance drops across all methods, SynCell remains robust (AUROC: 0.602, AUPRC: 0.369, F1: 0.336), demonstrating its ability to infer transferable biological interaction mechanisms beyond memorized drug identities. For all panels, left: ROC curves (FPR vs. TPR); middle: PR curves (Recall vs. Precision); right: bar charts with overlaid scatter points showing mean $\pm$ s.d. across five independent runs. Statistical significance ($p<0.05$) is denoted by asterisks compared to SynCell.
    }
    \label{fig:5-split}
\end{figure}

\begin{figure}[htbp]
    \centering
    \includegraphics[width=1\linewidth]{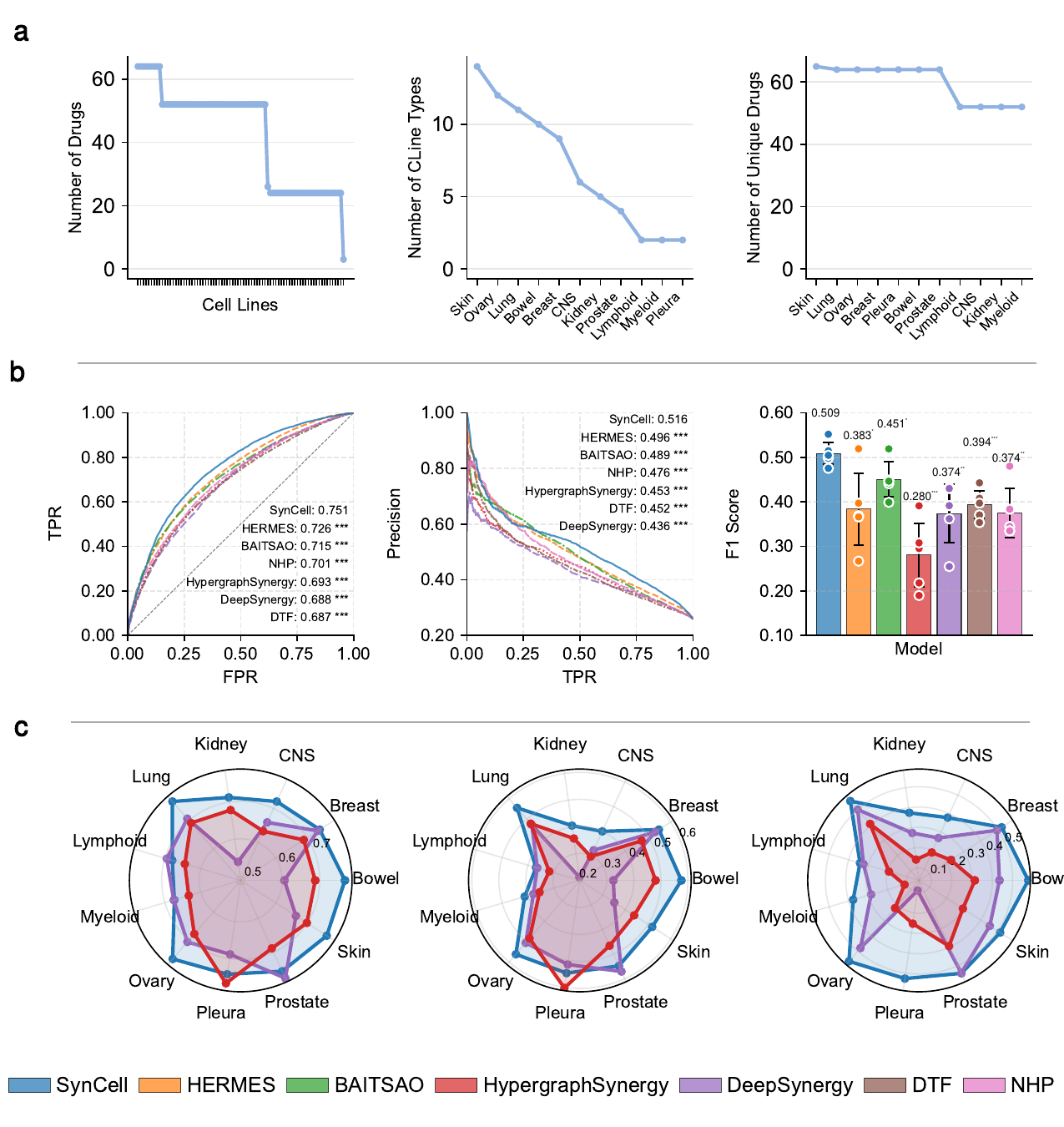}
    \caption{
    \textbf{Cell-line-specific molecular contexts drive generalization more effectively than coarse organ-level taxonomy.} 
    \textbf{a,} Data summaries: distribution of drugs per cell line (left), number of cell line types per organ category (middle), and number of unique drugs per organ (right), illustrating heterogeneity in biological coverage across tissue types. 
    \textbf{b,} Random cell-line split performance: SynCell achieves the highest AUROC (0.751), AUPRC (0.516) and F1 (0.509), indicating effective modeling of fine-grained cellular variation when transferred across diverse unseen backgrounds. 
    \textbf{c,} Organ-level performance comparison across eleven tissue categories: radar plots show AUROC (left), AUPRC (middle), and F1 (right). SynCell consistently outperforms all baselines, with notable advantages in epithelial tumor types (e.g., Ovary, Lung, Bowel, Skin), while maintaining robustness in biologically challenging organs (e.g., CNS, Lymphoid, Myeloid), where all methods exhibit performance degradation. Error bars represent 95\% confidence intervals across five random splits ($n=5$). Statistical significance ($p<0.001$) is denoted by asterisks.
    }
    \label{fig:organ_average}
\end{figure}

SynCell consistently outperformed state-of-the-art baselines across multiple benchmarks and independent validation settings, with the most pronounced gains observed under fully cold-start conditions in which both drugs were absent from training (\ref{fig:5-split}. 2c). Cell-line-specific protein–protein interaction networks proved critical for generalization, substantially outperforming models that rely solely on organ-level taxonomic labels (\ref{fig:organ_average}). Interpretability analyses further revealed that SynCell captures biologically meaningful mechanisms underlying drug synergy through mechanism-of-action profiling and pathway enrichment, supporting its utility for hypothesis generation and translational research (Fig. \ref{fig:Interpretability}).

To assess the predictive performance and generalizability of SynCell, we evaluated the framework across four complementary dimensions: (1) cross-dataset benchmark evaluation under various drug splitting strategies; (2) organ-stratified generalization analysis to assess model robustness across different disease categories; (3) extension to high-order drug combinations to test model scalability; (4) comprehensive interpretability analysis linking predictions to biological pathways. The following subsections present these results in detail, with corresponding figures illustrating our key findings.

\subsection{SynCell adapt well in unseen drug synergy prediction}

SynCell was evaluated across three progressively challenging generalization settings mirroring distinct stages of clinical drug development. Across all settings, SynCell consistently outperformed state-of-the-art baselines, inferring transferable biological mechanisms rather than memorizing drug identities. This addresses the critical clinical need to optimize existing regimens for drug-resistant cases where systematic approaches are required to overcome resistance or reduce toxicity~\cite{humphrey2011opportunities}.

The DrugComb setting (known drugs, novel combinations) addresses the clinical need to optimize existing regimens for drug-resistant cases. Physicians are familiar with drug safety profiles but require systematic approaches to identify synergistic combinations that overcome resistance or reduce toxicity, such as designing second-line therapies for patients failing standard treatment. Under this setting, SynCell achieves an AUROC of 0.7955 $\pm$ 0.0130, surpassing DeepSynergy (0.7619)~\cite{preuer2018deepsynergy} and BAITSAO (0.7302)~\cite{liu2025baitsao}. DeepSynergy relies on static molecular fingerprints processed by deep MLPs, while BAITSAO utilizes LLM-generated semantic embeddings; both struggle to capture dynamic biological interactions. SynCell's advantage stems from its contextualized PPI networks, which model how drug effects propagate through cell-specific signaling pathways. All improvements are statistically significant ($p < 0.05$), indicating SynCell effectively identifies synergistic patterns beyond simple feature co-occurrence.

The DrugSingle setting (one drug unseen) reflects the scenario of introducing new candidates to standard care, commonly encountered in clinical trial design or drug repurposing initiatives. SynCell maintains robust performance with an AUROC of 0.6460 $\pm$ 0.0254, outperforming HERMES (0.6335)~\cite{wu2025hermes} and NHP (0.5859)~\cite{yadati2020nhp}. HERMES employs gated hypergraph convolutions but lacks cell-line-specific molecular wiring, leading to performance degradation when drug structures deviate from training distributions. In contrast, SynCell leverages protein-level interactions to extrapolate pharmacological effects to novel entities. This suggests that modeling drug-protein-cell relations provides a more stable foundation for generalization than hypergraph co-occurrence patterns alone.

The DrugDouble setting (both drugs unseen) represents the most demanding evaluation, corresponding to early-stage discovery or rare disease scenarios with no prior synergy data. Despite substantial performance drops across all methods, SynCell achieves the highest AUROC of 0.6017 $\pm$ 0.0318, significantly exceeding HERMES (0.5639)\cite{wu2025hermes} and HypergraphSynergy (0.5568) with $p < 0.05$. While baseline models rely on learned drug embeddings that vanish in zero-shot settings, SynCell infers synergy through shared protein targets and pathway perturbations. These results confirm that SynCell generalizes beyond memorized drug identities, actively predicting synergy in realistic drug discovery settings where molecular data is sparse.

\subsection{SynCell generalizes to unseen cellular contexts}

To evaluate the clinical translational potential of SynCell, we designed two complementary generalization scenarios that mirror challenges in personalized oncology and drug repurposing, respectively. Cross-cell line generalization addresses intratumoral heterogeneity within the same cancer type. In clinical practice, even among patients with the same tissue origin, distinct molecular profiles lead to divergent drug responses. This setting simulates the model's ability to generalize from known patients, represented by training cell lines, to unknown patients, represented by test cell lines, with the core value of supporting molecular feature-based personalized combination recommendations rather than relying on coarse organ taxonomy. Furthermore, this setting encompasses scenarios involving rare subtypes and tumor evolution from primary site to resistant recurrence, aiming to reduce experimental trial-and-error costs for data-scarce orphan cancers and provide computational basis for dynamic treatment strategies post-resistance. Cross-organ generalization corresponds to drug repurposing and rare cancer treatment scenarios. Many rare cancers lack sufficient training data, whereas common cancers are data-rich. This setting evaluates whether the model can transfer knowledge from data-rich cancer types to data-scarce indications, aligning with global efforts to accelerate drug discovery for the diverse landscape of cancer types\cite{bray2024cancer}. Its practical value lies in accelerating drug discovery for rare cancers and validating the tissue-agnostic nature of synergistic mechanisms. If the model demonstrates robust generalization across multiple organs, it supports the design of cross-cancer basket trials, optimizing clinical resource allocation.

We first assessed performance under the Random CLine split, where cell lines are randomly partitioned between training and testing sets while preserving drug identity overlap. This setting examines the model's ability to generalize to unseen cellular backgrounds, analogous to predicting responses in new patient-derived cell lines. SynCell achieves the highest AUROC of 0.751 and AUPRC of 0.516, outperforming all baselines with statistical significance. Notably, SynCell surpasses HERMES, a state-of-the-art hypergraph model that captures higher-order drug interactions but lacks explicit protein-level context~\cite{wu2025hermes}, by 2.5 percentage points in AUROC. It also exceeds BAITSAO, which relies on LLM-generated drug features~\cite{liu2025baitsao}, suggesting that structural biological priors, specifically cell-specific PPIs, provide more transferable signals than semantic drug descriptions alone. Other baselines, including DeepSynergy, DTF~\cite{sun2020dtf}, and HypergraphSynergy, show clear performance gaps. These results demonstrate that SynCell maintains strong predictive power when transferred across diverse cellular contexts, indicating effective modeling of cell line-specific biological variation to support personalized treatment selection.

To further investigate robustness at the tissue level, we evaluated performance across eleven organ categories. Averaged across organs, SynCell achieves the highest mean AUROC of 0.7714 and AUPRC of 0.5441, consistently outperforming competing methods. However, performance varies by tissue type, reflecting intrinsic biological complexity rather than model instability. For example, SynCell achieves strong performance in epithelial tumors such as Lung and Bowel. In biologically challenging categories like CNS/Brain and Myeloid, where AUPRC values across all models drop to lower ranges, SynCell still attains the highest AUROC, outperforming HERMES and BAITSAO. Although isolated cases exist where another method achieves slightly higher metrics, SynCell provides more balanced performance across AUROC, AUPRC, and F1. Collectively, these analyses indicate that while organ-level taxonomy provides a coarse grouping, cell-line-specific molecular contexts are more predictive of synergy than tissue-of-origin, validating SynCell's design principle of contextualized PPI networks.

\subsection{SynCell scales robustly to high-order drug combinations}
\begin{figure}[htbp]
    \centering
    \includegraphics[width=1\linewidth]{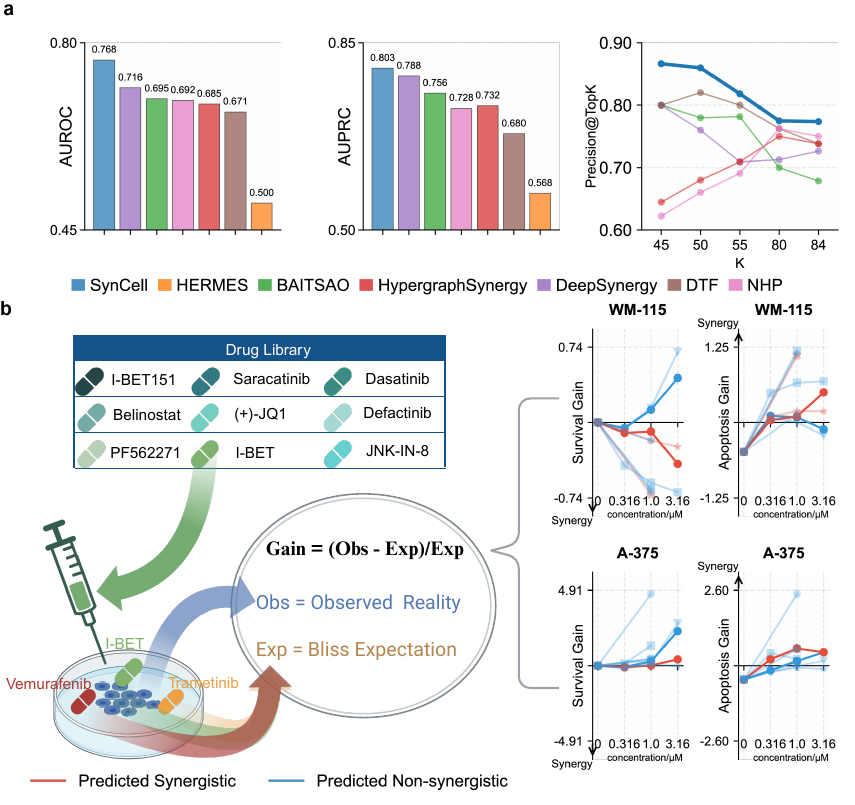}
    \caption{
    \textbf{SynCell scales robustly to three-drug combinations, capturing context-dependent synergy patterns in structured discrete dose grids.} 
    \textbf{a,} Model performance on triple-drug prediction. Left: AUROC comparison showing SynCell achieves the best performance (0.768), outperforming DeepSynergy (0.716), BAITSAO (0.695), NHP (0.692), HypergraphSynergy (0.685), and DTF (0.671), with HERMES substantially lower (0.500). Middle: AUPRC comparison where SynCell again leads (0.803), followed by DeepSynergy (0.788) and BAITSAO (0.756). Right: Precision@TopK curves across varying K, where SynCell consistently maintains the highest precision, demonstrating superior ranking quality in high-confidence regions. 
    \textbf{b,} Experimental validation workflow and representative case studies. Left: drug library and experimental design for three-drug combination screening, with synergy quantified as Gain = (Obs $-$ Exp)/Exp under the Bliss independence model. Right: representative survival gain and apoptosis gain curves for WM-115 and A-375 cell lines across increasing concentrations of the third drug. Predicted synergistic combinations (red) exhibit stronger-than-expected viability reduction and apoptosis induction, whereas predicted non-synergistic combinations (blue) closely follow or underperform the Bliss expectation baseline. These results demonstrate that SynCell captures context-dependent higher-order drug interactions and produces biologically consistent ranking signals aligned with experimental outcomes.
    }
    \label{fig:high_order}
\end{figure}

To further evaluate model scalability beyond pairwise interactions, we extended the prediction task to three-drug combinations using the triple-drug subset from DrugCombDB\cite{liu2020drugcombdb}. In this dataset, Drug1 and Drug2 are administered according to a predefined five-point joint concentration schedule, where the paired doses are fixed at (0, 0), (0.1, 0.01), (0.316, 0.0316), (1, 0.1), and (3.16, 0.316) $\mu$M. In contrast, Drug3 is evaluated independently at four discrete concentration levels: 0, 0.316, 1.0, and 3.16 $\mu$M. Consequently, each triple-drug combination is measured over a structured discrete dose grid defined by the fixed Drug1–Drug2 joint schedule and the varying Drug3 concentrations, rather than over a continuous concentration space. This experimental design preserves biologically meaningful dosage regimes while substantially increasing the combinatorial interaction complexity compared to pairwise settings.

Under this high-order setting, SynCell achieves an AUROC of 0.768 and an AUPRC of 0.803 (Figure~\ref{fig:high_order}(a)), outperforming all baselines. Notably, SynCell surpasses the second-best method, DeepSynergy (AUROC: 0.716, AUPRC: 0.788), by a clear margin. In terms of ranking quality, SynCell also maintains superior Precision@TopK across varying K, indicating improved identification of highly synergistic combinations. These results demonstrate that SynCell scales effectively to higher-order interaction modeling without degradation despite the exponential expansion of the combinatorial space.

\textbf{Cell line–specific analysis of predicted top and bottom synergies.}

To assess whether the model captures biologically meaningful interaction patterns, we analyzed representative top- and bottom-ranked Drug3 combinations predicted by SynCell in two melanoma cell lines (WM-115 and A-375). Notably, the identities of high- and low-ranked Drug3 differ across cell lines, suggesting that SynCell captures context-dependent synergy rather than relying on globally dominant drug effects.

Figure~\ref{fig:high_order}(b) shows the experimentally observed survival gain and apoptosis gain across increasing Drug3 concentrations (0.316, 1.0, and 3.16 $\mu$M), while Drug1 and Drug2 follow their fixed joint dose schedule. Synergy is quantified as Gain = (Obs $-$ Exp)/Exp relative to the Bliss independence expectation.

For the top-ranked combinations (red curves), increasing Drug3 concentration leads to progressively stronger negative survival gain and positive apoptosis gain, indicating supra-additive effects. This trend is particularly evident in WM-115, where higher concentrations induce substantial apoptosis gain.

In contrast, bottom-ranked combinations (blue curves) exhibit limited deviation from the additive expectation. Survival and apoptosis gains remain close to zero or show weaker trends across concentrations, indicating minimal synergistic interaction.

These results confirm that SynCell’s ranking aligns with experimentally observed synergy strength and that the model effectively distinguishes strong and weak three-drug interactions under structured discrete dose settings.

\subsection{SynCell identifies biologically meaningful drug interaction patterns through MoA-level analysis and pathway enrichment}

\begin{figure}[htbp]
    \centering
    \includegraphics[width=0.8\linewidth]{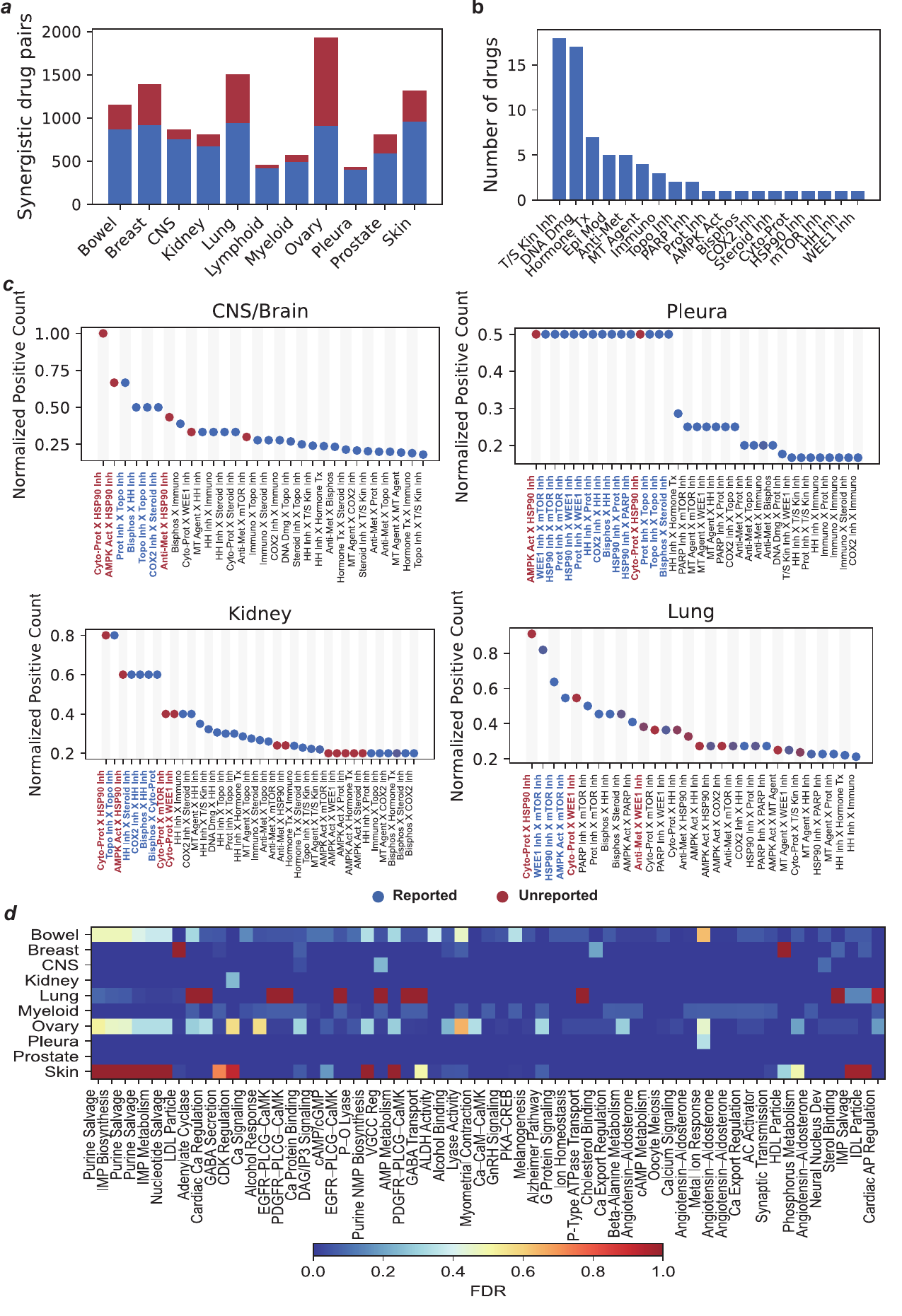}
    \caption{
    \textbf{SynCell uncovers biologically meaningful drug interaction patterns through mechanism-of-action–level analysis and pathway enrichment.} 
    \textbf{a,} Number of synergistic drug pairs across organ types, separated into previously reported (blue) and newly predicted (red) interactions. SynCell substantially expands the landscape of candidate synergistic combinations beyond known data, particularly in Ovary, Lung, and Breast tissues. 
    \textbf{b,} Distribution of drugs across mechanism-of-action (MoA) categories, illustrating the diversity and imbalance of pharmacological classes in the dataset. 
    \textbf{c,} Organ-specific MoA combination analysis. For each organ, the top-ranked MoA combinations are shown based on normalized synergy density (number of positive pairs divided by total possible pairs). Blue points denote reported synergies, while red points indicate novel predictions. SynCell prioritizes biologically plausible and context-dependent MoA interactions, with distinct patterns observed across organs. 
    \textbf{d,} Pathway enrichment analysis based on attention-prioritized proteins. Heatmap shows significantly enriched pathways across organ types, with color indicating FDR-adjusted significance. Enriched pathways reveal context-specific biological mechanisms underlying predicted drug synergy.
}    \label{fig:Interpretability}
\end{figure}

To investigate whether SynCell captures biologically meaningful interaction mechanisms, we analyzed predicted drug synergies at the level of mechanism-of-action (MoA) combinations and linked these patterns to underlying signaling pathways.

Starting from the drug and cell line space in the dataset, we constructed the full combinatorial space of possible drug pairs within each cell-line context. Predicted positive pairs from SynCell were combined with experimentally reported synergistic pairs to form an expanded interaction set. Each drug was mapped to its MoA category, enabling analysis at the level of MoA–MoA combinations rather than individual compounds.

Figure~\ref{fig:Interpretability}(a) summarizes the number of synergistic drug pairs across organ types, distinguishing between reported and newly predicted interactions. Across multiple tissues, SynCell identifies a substantial number of previously unreported synergistic pairs, effectively expanding the known interaction landscape. This expansion is particularly pronounced in Ovary, Lung, and Breast tissues, suggesting that the model generalizes beyond memorized training pairs.

To understand the underlying pharmacological space, Figure~\ref{fig:Interpretability}(b) shows the distribution of drugs across MoA categories. The dataset exhibits a highly imbalanced structure, with certain MoA classes (e.g., kinase inhibitors and DNA-targeting agents) dominating the space, which further motivates analyzing interactions at the MoA level.

We next examined organ-specific patterns of MoA–MoA interactions. For each organ, we ranked MoA combinations based on normalized synergy density (defined as the number of positive pairs divided by the total number of possible pairs under that combination). Figure~\ref{fig:Interpretability}(c) presents the top-ranked MoA combinations, with reported interactions shown in blue and newly predicted interactions in red. Notably, the highest-ranked MoA combinations differ substantially across organs, indicating strong context dependence. In several cases, SynCell prioritizes novel MoA combinations (red) that achieve comparable or higher normalized synergy density than reported ones, suggesting that the model identifies biologically plausible yet previously unexplored interaction mechanisms.

Notably, among the top-ranked novel predictions, SynCell identified the combination of PARP inhibitors (e.g., talazoparib) and WEE1 inhibitors (e.g., adavosertib) as highly synergistic, particularly in ovarian and breast cancer contexts. This prediction aligns with emerging clinical evidence demonstrating that co-inhibition of PARP and WEE1 induces synthetic lethality in homologous recombination-proficient tumors by exacerbating replication stress and forcing premature mitotic entry~\cite{palve2021noncanonical,smith2024atr,fang2019sequential}. Specifically, while PARP inhibition stalls replication forks, WEE1 inhibition abrogates the G2/M checkpoint and impairs homologous recombination repair, leading to replication catastrophe~\cite{smith2024atr}. The model's ability to prioritize this mechanistically grounded combination without explicit training on these specific pairwise interactions underscores its capacity to infer higher-order biological dependencies from the contextualized PPI network.

Complementing these findings, SynCell also prioritized combinations of DNA damaging agents (e.g., cisplatin) and topoisomerase inhibitors (e.g., topotecan), predicting strong synergy driven by replication fork collapse. This aligns with established mechanisms where topoisomerase inhibitors stabilize Top1-DNA complexes, while DNA damaging agents exacerbate replication stress, leading to irreversible double-strand breaks when checkpoint pathways (ATR/CHK1) or homologous recombination (RAD51) are co-inhibited~\cite{jo2021novel,mann2020combined,urrutia2021combined}. Specifically, blocking checkpoint signaling prevents replication fork stabilization, forcing cells with topoisomerase-induced damage into mitotic catastrophe~\cite{jo2021novel,tiwari2021triplex}. Furthermore, synergy is enhanced in contexts with deficient DNA repair machinery, such as NER defects, highlighting the model's sensitivity to molecular context~\cite{tiwari2021triplex,hirai2022small}. The ability of SynCell to capture these complex DNA damage response (DDR) interactions without explicit training on these specific pairwise rules underscores its mechanistic reasoning capabilities.

To further connect these interaction patterns to underlying biology, we performed pathway enrichment analysis using proteins prioritized by SynCell's attention mechanism. Specifically, for each organ, proteins were ranked by their contribution to predicted synergy, and the top-ranked subset was subjected to over-representation analysis (ORA) using the hypergeometric test \cite{reimand2019}. Pathway annotations were sourced from canonical databases including KEGG and Reactome \cite{kegg2023, fabregat2018}. As shown in Figure~\ref{fig:Interpretability}(d), the resulting pathways exhibit clear tissue-specific patterns after Benjamini-Hochberg correction (FDR < 0.05). For example, signaling pathways related to receptor tyrosine kinases (RTK), MAPK cascades, and metabolic regulation are selectively enriched in specific organs, reflecting known biological heterogeneity across tissue types \cite{menche2015}. 

Notably, the PDGFR–PLCG–CaMK signaling pathway was significantly enriched in lung tissue predictions. This pathway has been extensively implicated in non-small cell lung cancer (NSCLC), where PDGF/PDGFR signaling promotes tumor stroma proliferation via paracrine mechanisms, facilitates epithelial-mesenchymal transition, and correlates with lymphatic metastasis and poor prognosis \cite{pdgf_nsclc_2014}. The selective enrichment of this pathway in lung-related predictions, despite its absence from the training labels for many drug combinations, demonstrates SynCell's capacity to recover biologically grounded, context-specific mechanistic insights beyond generic drug-effect patterns.

This organ-specific enrichment aligns with established oncogenic drivers, suggesting that SynCell successfully captures context-dependent signaling vulnerabilities rather than generic drug effects.

Together, these results demonstrate that SynCell does not rely on superficial correlations between drug identities, but instead captures higher-level pharmacological interaction structures and links them to biologically meaningful, context-specific signaling mechanisms. This provides strong evidence that the model learns interpretable and mechanistically grounded representations of drug synergy, facilitating hypothesis generation for translational research \cite{zitnik2019}. 
\section{Methods}

\label{sec:methods}

\subsection{Problem Definition}
\label{sec:problem_definition}

We formalize cell-context-aware drug synergy prediction as a link prediction task on a dynamic, heterogeneous knowledge graph. Formally, let $\mathcal{G} = (\mathcal{V}, \mathcal{E})$ denote a heterogeneous knowledge graph where nodes $\mathcal{V}$ represent biological entities across four types: drugs ($\mathcal{V}_D$), proteins ($\mathcal{V}_P$), cell lines ($\mathcal{V}_C$), and diseases ($\mathcal{V}_{Dis}$). The edge set $\mathcal{E}$ encodes biological relationships, including drug-target interactions, protein-protein interactions (PPIs), cell-line-specific protein expression, and disease-protein associations. Unlike conventional approaches that operate on static PPI networks, we define the graph structure as context-dependent. For each cell line $c_k \in \mathcal{V}_C$, we derive a contextualized subgraph $\mathcal{G}_k \subseteq \mathcal{G}$ by masking global PPI edges based on the protein expression profile of $c_k$, ensuring that information propagation occurs only through biologically active pathways specific to that cellular environment.

Given a drug pair $(d_i, d_j)$ and a specific cell line $c_k$, the objective is to predict the synergy score $y_{ijk} \in \{0, 1\}$, indicating whether the combination exhibits synergistic effects in the molecular context of $c_k$. This formulation extends standard pairwise inference to a multi-relational, zero-shot challenge where the model must generalize to unseen drug combinations or novel cell lines by leveraging the underlying biological topology. Our approach induces inductive priors by incorporating factual knowledge from the knowledge graph into the model, enhancing its reasoning capabilities for hypothesis formation and synergy prediction in data-sparse scenarios.

\subsection{Overview of the SynCell Framework}
\label{sec:framework_overview}

SynCell is a deep-learning framework designed for mechanistic synergy prediction based on cell-line-specific molecular networks. To address the heterogeneity of cellular contexts, SynCell is composed of four integrated modules: (1) a \textbf{Context-Aware Graph Construction} module that dynamically derives cell-line-specific PPI subnetworks from global interactomes; (2) a \textbf{Heterogeneous Graph Encoder} that learns biologically meaningful network representations for each entity through relation-specific message passing; (3) a \textbf{Contextual Feature Modulation} mechanism that adaptively transforms drug embeddings based on the cellular environment; and (4) a \textbf{Multi-Task Prediction and Interpretability} module that jointly optimizes synergy prediction and organ classification while generating mechanistic explanations.

The framework unfolds through three synergistic phases designed to enhance generalization and biological interpretability (Fig.~\ref{fig:framework}c). First, in the \textbf{graph construction phase}, we integrate multimodal data to build cell-line-specific PPI subnetworks, prioritizing fine-grained molecular contexts over coarse organ-level taxonomy. This design is motivated by our finding that cell-line-specific wiring drives generalization more effectively than tissue-of-origin labels. Second, in the \textbf{representation learning phase}, the model employs a dual-pathway architecture. A primary pathway propagates information via relation-specific convolutions, while a secondary hypergraph module captures higher-order interactions among drug pairs and diseases. Crucially, we introduce a context-aware feature modulation layer that generates scaling and shifting parameters conditioned on the cell-line embedding to dynamically adapt drug representations. This ensures that the same drug receives distinct embeddings in different cellular environments, reflecting context-dependent pharmacological effects. Third, in the \textbf{prediction and validation phase}, the modulated features are fed into parallel heads for synergy classification and organ prediction. We further implement an attention-based perturbation framework to identify key protein drivers, establishing protocols to corroborate computational predictions with clinical and experimental evidence. This comprehensive formulation treats synergy prediction not merely as a classification task, but as a mechanistic inference problem grounded in cellular context.

\subsection{Data Curation and Ethics Statement}

Most data used for this study were obtained from publicly available knowledge repositories. All computational data were deidentified and anonymized prior to analysis. We curated a multimodal biomedical dataset from four integrated sources to establish the foundation for SynCell. As delineated in Figure~\ref{fig:framework}b, drug synergy measurements were obtained from DrugCombDB~\cite{liu2020drugcombdb}, comprising 72,364 combination samples across 124 anti-cancer compounds and 48 human cancer cell lines. Gene expression profiles were sourced from the Cancer Dependency Map (DepMap)~\cite{oneil2016oncology}, containing transcriptomic data for 1,799 cell lines across 19,221 genes. Comprehensive biological relationships were extracted from PrimeKG~\cite{chandak2022primekg}, including 51,306 drug-target interactions, 413 disease entities with associated pathological descriptions, and disease-protein associations. Molecular structures of drugs in SMILES format were retrieved from DrugBank. 

Data integration required extensive cross-referencing to resolve identifier mismatches between databases, particularly for protein targets and cell line annotations. We applied rigorous quality control protocols, removing entries with missing values or inconsistent annotations. Expression data underwent quantile normalization to minimize batch effects across different profiling platforms. The resulting unified dataset preserves biological fidelity while enabling coherent graph construction. 

\subsection{Problem Definition and Heterogeneous Graph Construction}

\subsubsection*{Graph Curation}
The heterogeneous graph incorporates four biological entity types: drug nodes representing 124 compounds from DrugCombDB, cell line nodes corresponding to 55 cancer cell lines with complete transcriptomic profiles, protein nodes comprising 1,257 key proteins selected through integrated topological-biological scoring, and disease nodes encompassing 413 cancer-related pathologies from PrimeKG. Protein selection employed a composite scoring system combining PageRank centrality in the STRING-derived PPI network~\cite{szklarczyk2023string}, degree centrality, and pan-cancer expression coverage, retaining top-ranked proteins with documented therapeutic relevance.

We define five biologically grounded relation types to construct the graph topology. 
\begin{enumerate}
    \item \textbf{Cell line-protein expression:} Edges were established using DepMap expression data, where proteins with expression exceeding the 80th percentile across all DepMap cell lines were connected to their respective cell lines, generating 12,874 high-confidence expression edges that capture context-specific proteome activity. \textbf{This threshold was selected based on optimization on the validation set to balance network sparsity and biological coverage, consistent with strategies for constructing tissue-specific interactomes~\cite{greene2015understanding}.}
    \item \textbf{Drug-protein targeting:} Relations were extracted directly from PrimeKG and filtered to retain only interactions involving our drug set, yielding 892 high-confidence edges after manual curation.
    \item \textbf{Protein-protein interaction:} Edges were derived from STRING with confidence scores greater than 400, restricted to our protein set to form 16,591 context-agnostic interactions. Crucially, these global interactions were subsequently contextualized to create cell line-specific PPI subnetworks.
    \item \textbf{Cell line similarity:} Edges connected cell lines based on transcriptomic similarity with Pearson correlation greater than 0.85 on 651 cancer driver genes, forming a cell line similarity backbone with 328 edges.
    \item \textbf{Disease-protein association:} Relations were sourced from PrimeKG's disease-protein associations, incorporating 2,147 edges linking pathologies to molecular mechanisms.
\end{enumerate}
This construction yields a heterogeneous graph structure that explicitly encodes the molecular context required for zero-shot generalization to unseen cell lines and drug combinations.

\subsection{Multimodal Feature Encoding and Initialization}

To ensure meaningful representation learning across diverse biological entities, we derive initial feature representations through specialized encoders tailored to each entity's data modality. Formally, for each node $i \in V$ in the heterogeneous graph, we initialize a latent representation denoted as $h_i^{(0)}$. Drug features were generated from SMILES strings using ChemBERTa (77 million parameters), where tokenized sequences passed through a 12-layer Transformer encoder followed by mean pooling of final hidden states, producing 384-dimensional vectors encoding structural and functional properties. Cell line and protein features were generated via truncated Singular Value Decomposition (SVD) of the cell line-protein expression matrix, producing 1024-dimensional embeddings capturing co-expression patterns across cellular contexts. Disease features leveraged PubMedBERT to process textual descriptions and names from PrimeKG, generating 768-dimensional semantic embeddings that capture pathological characteristics. To stabilize training dynamics and align modalities, all features underwent layer normalization before being linearly projected to a shared 256-dimensional latent space:
\begin{equation}
    h_i^{(0)} = \text{LayerNorm}(W_{\text{proj}} \cdot \phi_{\text{modality}}(x_i)),
\end{equation}
where $\phi_{\text{modality}}$ represents the modality-specific encoder and $W_{\text{proj}}$ is a learnable projection matrix. This unified initialization ensures that subsequent graph propagation operates within a consistent geometric space.

\subsection{Context-Aware Heterogeneous Graph Learning}

Our architecture processes the heterogeneous graph through dual complementary propagation pathways that converge for context-aware prediction. The model unfolds through three synergistic phases: heterogeneous message passing, contextual modulation, and multi-task prediction.

\textbf{Heterogeneous message passing.} The primary pathway employs a two-layer heterogeneous graph convolutional network using relation-specific SAGEConv layers~\cite{hamilton2017inductive}. For every relationship type $r \in \mathcal{R}$, we calculate a transformation of node embedding from the previous layer $h_i^{(l-1)}$ by applying a relationship-specific weight matrix $W_r^{(l)}$:
\begin{equation}
    m_{r,i}^{(l)} = W_r^{(l)} h_i^{(l-1)}.
\end{equation}
For each node $i$, we aggregate incoming messages from neighboring nodes of each relation $r$, denoted as $\mathcal{N}_r(i)$, by taking the average of these messages:
\begin{equation}
    \tilde{m}_{r,i}^{(l)} = \frac{1}{|\mathcal{N}_r(i)|} \sum_{j \in \mathcal{N}_r(i)} m_{r,j}^{(l)}.
\end{equation}
We then combine the node embedding from the last layer and the aggregated messages from all relationships to obtain the new node embedding:
\begin{equation}
    h_i^{(l)} = h_i^{(l-1)} + \sum_{r \in \mathcal{R}} \tilde{m}_{r,i}^{(l)}.
\end{equation}
All edge weights undergo log-normalization to balance their influence across different relation types. The secondary pathway utilizes a hypergraph module modeling higher-order relationships among drug pairs, cell lines, and associated diseases through hyperedge constructions that connect synergistic triplets.

\textbf{Contextual modulation.} To dynamically identify proteins most relevant to synergy prediction in specific cellular contexts, we introduce a Drug-Protein-Cell Cross-Attention mechanism. For each cell line, we retrieve its expressed protein set and compute attention weights between the drug-pair-cell triplet and each expressed protein. The attention module employs multi-head attention with temperature scaling, where query vectors represent concatenated drug and cell line features, while key and value vectors correspond to protein features. The context-aware protein representation is computed as a weighted sum of protein features according to these attention weights. Subsequently, the propagated features from both pathways are fused and processed through a \textbf{context-conditioned affine transformation layer} that dynamically adapts representations to cellular contexts. Specifically, for a drug pair and cell line triplet, \textbf{learnable scaling ($\gamma$) and shifting ($\beta$) parameters} are generated by a context encoder (a multilayer perceptron) that processes concatenated drug, cell line, and aggregated disease features. The \textbf{contextualized features} $\hat{h}$ are computed as:
\begin{equation}
    \hat{h} = \gamma \odot h + \beta,
\end{equation}
This operation \textbf{injects cellular context directly into the drug representation space}, ensuring that the same drug receives distinct embeddings in different cellular environments to reflect context-dependent pharmacological effects.

\textbf{Multi-task prediction.} The modulated features feed into two parallel prediction heads: a synergy classifier using a symmetric multilayer perceptron architecture invariant to drug order, and an organ classifier predicting the tissue-of-origin from eleven cancer types. The joint loss function combines both objectives to enhance generalization across unseen biological environments:
\begin{equation}
    \mathcal{L} = \mathcal{L}_{\text{synergy}} + \lambda \mathcal{L}_{\text{organ}},
\end{equation}
where $\mathcal{L}_{\text{synergy}}$ is the binary cross-entropy loss for synergy prediction, $\mathcal{L}_{\text{organ}}$ is the cross-entropy loss for organ classification, and $\lambda=0.2$ is a weighting factor balanced based on cross-validation performance. This formulation treats synergy prediction not merely as a classification task, but as a mechanistic inference problem grounded in cellular context.

\subsection{Interpretability Analysis Framework}

To ensure the clinical translatability of SynCell, we developed a comprehensive interpretability framework designed to validate model predictions against biological knowledge and identify key protein drivers underlying synergy mechanisms. Unlike black-box approaches, SynCell leverages its attention-based architecture to provide mechanistic insights. \textbf{Our framework aligns with established graph explainability standards, incorporating principles from perturbation-based methods such as GNNExplainer \cite{ying2019gnnexplainer} and GraphMask \cite{schlichtkrull2021interpreting} to ensure faithfulness and stability.} For each predicted drug--cell line synergy score, we extract attention weights from the Drug--Protein--Cell Cross-Attention module. These weights quantify the contribution of each expressed protein to the final prediction. We define the importance score $I_p$ for protein $p$ as the normalized attention weight averaged across attention heads. To identify critical mediators, we select the top 5\% of proteins ranked by $I_p$. We chose this discrete subset approach over ranked-list methods (e.g., GSEA) because attention weights naturally highlight a sparse set of high-confidence drivers suitable for over-representation analysis \cite{reimand2019}.

To validate the biological relevance of these identified proteins, we perform \textit{in silico} graph perturbation experiments, analogous to faithfulness tests in explainable AI \cite{schlichtkrull2021}. Specifically, we systematically remove the top-ranked proteins and all their associated edges from the knowledge graph, creating a perturbed graph $\mathcal{G}'$. We then recompute synergy predictions using $\mathcal{G}'$ and measure the prediction shift $\Delta y = |y_{\mathcal{G}} - y_{\mathcal{G}'}|$. A significant $\Delta y$ indicates that the removed proteins are causally important for the model's decision-making process. As a control, we perform identical perturbation experiments with randomly selected protein sets to establish baseline expectations. 

For pathway enrichment, we employed the hypergeometric test to assess the over-representation of annotated pathways within the selected protein subset. The background set was defined as all 1,257 proteins present in the knowledge graph to ensure consistency with the model's input space. P-values were adjusted for multiple testing using the Benjamini-Hochberg procedure \cite{benjamini1995}, with pathways considered significantly enriched at an FDR < 0.05. This framework not only provides biological validation of model decisions but also enables the identification of potentially novel protein mediators of drug synergy, facilitating hypothesis generation for wet-lab validation.

\subsection{Training Strategy and Optimization}

The model is trained using a unified optimization protocol designed to balance synergy prediction accuracy with biological consistency. The primary objective is to minimize the binary cross-entropy loss between predicted synergy scores and ground-truth labels. To address the inherent class imbalance in synergy data (approximately 12\% positive samples), we implement a positive-sample upsampling strategy during batch construction to achieve a balanced 1:1 positive-to-negative ratio. Additionally, SynCell employs a multi-task learning framework, jointly optimizing synergy prediction and organ classification. The total loss function $\mathcal{L}_{total}$ is defined as:
\begin{equation}
    \mathcal{L}_{total} = \mathcal{L}_{synergy} + \lambda \mathcal{L}_{organ}
\end{equation}
where $\lambda=0.2$ is a weighting factor determined via cross-validation to balance the auxiliary task without dominating the primary objective.

Hyperparameters are systematically tuned using the Optuna framework with Bayesian search across multiple trials. Each trial evaluates model configurations through 5-fold cross-validation on the training set, with the Area Under the Precision-Recall Curve (AUPRC) serving as the primary optimization metric due to its robustness against class imbalance. The search space encompasses learning rates sampled log-uniformly between $1\times10^{-5}$ and $1\times10^{-3}$, hidden dimensions selected from $\{64, 128, 256\}$, attention heads from $\{2, 4, 8\}$, dropout rates from 0.05 to 0.3, and temperature parameters from 0.5 to 2.0. Optimization employs the Adam optimizer coupled with dynamic learning rate scheduling and early stopping based on validation AUROC with a patience of 10 epochs. All experimental results report mean and standard deviation over five independent runs with fixed random seeds to ensure statistical reliability and reproducibility.

\subsection{Evaluation Protocols and Experimental Setup}
\label{sec:evaluation}

To rigorously assess generalization capabilities, we implement five progressively challenging data splitting strategies that mirror distinct stages of clinical drug development. 
\begin{itemize}
    \item \textbf{Cross-Cell Line (CLine) Split:} Stratifies samples by cell line, ensuring all cell lines in the test set are completely unseen during training. This evaluates the model's ability to generalize across diverse cellular contexts.
    \item \textbf{Drug Combination (DrugComb) Split:} Partitions data based on drug pairs, with test combinations entirely absent from training, simulating the optimization of existing regimens.
    \item \textbf{Single Drug (DrugSingle) Split:} A semi-cold-start scenario where at least one drug in each test pair is unseen, reflecting drug repurposing or add-on therapy scenarios.
    \item \textbf{Double Drug (DrugDouble) Split:} The most stringent zero-shot scenario where both drugs in test pairs are absent from training, corresponding to early-stage discovery with sparse molecular data.
\end{itemize}
Crucially, in cell line and drug splits, disease associations are preserved only for training entities to prevent information leakage, \textbf{adhering to strict evaluation protocols recommended for graph-based machine learning to avoid optimistic bias~\cite{kaufman2012leakage, shchur2021pitfalls}.}

We conduct comprehensive comparative benchmarking against multiple baseline approaches to establish performance advantages. Baselines include a feature-only multilayer perceptron (ablation baseline), HERMES (state-of-the-art hypergraph model)~\cite{wu2025hermes}, DeepSynergy (established deep learning baseline)~\cite{preuer2018deepsynergy}, DeepDDS (graph neural network with attention)~\cite{wang2021deepdds}, GraphSynergy (network-inspired deep learning)~\cite{yang2021graphsynergy}, KGANSynergy (knowledge graph attention network)~\cite{zhang2023kgansynergy}, BAITSAO (recent multi-task learning approach)~\cite{liu2025baitsao}, NHP (neural hypergraph link prediction)~\cite{yadati2020nhp}, and DTF (deep tensor factorization)~\cite{sun2020dtf}. Primary synergy prediction is evaluated using Area Under the Receiver Operating Characteristic Curve (AUROC) and Area Under the Precision-Recall Curve (AUPRC), the latter serving as the primary optimization metric due to its robustness against class imbalance~\cite{ruopp2008youden}. Organ classification performance is assessed using weighted F1-score. Statistical significance of performance differences is determined through paired t-tests with significance level $\alpha=0.01$. We additionally compute calibration curves and confusion matrices to diagnose systematic biases, and measure computational efficiency via training throughput and inference latency. Ablation studies systematically evaluate individual components, including the hyperedge module, attention mechanisms, contextualization, and disease integration.

\section{Discussion}

We have developed SynCell, a heterogeneous graph neural network framework for contextualized drug synergy prediction. Our approach demonstrates superior performance across diverse evaluation scenarios, particularly in pharmacologically challenging settings. Here we discuss the implications of our findings, compare our methodology with existing approaches, acknowledge limitations, and outline future research directions.

\subsection{Interpretation of Key Findings}
SynCell achieves robust performance in pharmacologically challenging scenarios, particularly DrugSingle (64.60\% AUROC) and DrugDouble (60.17\% AUROC), representing a substantial advancement in predicting synergy for novel compounds under zero-shot conditions. This capability stems from our framework's effective utilization of protein-protein interaction networks as biological priors that encode functional relationships between drug targets, addressing the critical need for therapies in data-sparse contexts \cite{humphrey2011challenges, huang2024txgnn}. When drug-specific information is limited or absent, SynCell leverages the relational structure among proteins to infer mechanistic relationships, enabling informed predictions for novel chemical entities. The edge sampling study demonstrates remarkable robustness to sparse positive edge connectivity. In the DrugDouble scenario, performance with only 20\% of positive edges (60.95\% AUROC) slightly exceeded the full model (60.17\% AUROC), indicating that our framework effectively identifies and utilizes the most informative positive interactions. This finding has important practical implications for real-world applications where complete positive edge information may be unavailable or costly to obtain, as the model maintains strong performance even with substantially reduced edge information. The varying sensitivity to edge removal across evaluation strategies provides important insights. The Random split showed the greatest performance degradation with edge removal, likely because complete graph information offers incremental benefits when abundant training examples are available. In contrast, the DrugDouble scenario maintained robust performance even with substantial edge removal, suggesting that the essential biological priors for novel drug prediction are encoded within a core subset of protein interactions.

\subsection{Comparison with Existing Methods}
Our comparative analysis reveals distinct methodological strengths across evaluation scenarios. While HERMES demonstrates competitive performance in certain settings, SynCell exhibits advantages in pharmacologically novel scenarios \cite{wu2025hermes}. The 1.25 percentage point improvement in DrugSingle (64.60\% vs 63.35\%) and 3.78 percentage point improvement in DrugDouble (60.17\% vs 56.39\%) highlight the advantage of heterogeneous graphs over hypergraph representations for generalizing to unseen drugs under zero-shot conditions. This performance disparity suggests that explicit modeling of biological entities (proteins) and their relationships provides more transferable knowledge compared to hypergraph approaches that primarily capture co-occurrence patterns, aligning with recent systematic reviews on GNN efficacy in drug discovery \cite{besharatifard2024gnn}. The superiority in DrugComb scenarios (79.55\% AUROC vs 76.19\% for DeepSynergy and 73.02\% for BAITSAO) further validates that heterogeneous graphs offer more natural representations for capturing complex drug-drug interactions through shared biological mechanisms \cite{preuer2018deepsynergy, wang2024clinical}. Notably, DeepSynergy maintains relatively stable performance in novel drug scenarios (76.19\% in DrugComb), though lower than SynCell (79.55\%), indicating that feature-based methods provide baseline generalization while graph-based approaches enable more sophisticated relational reasoning through biological networks. The scalability of SynCell to higher-order drug combinations further demonstrates its versatility. In three-drug combination prediction, SynCell achieves an AUROC of 76.8\% and AUPRC of 80.3\%, substantially outperforming all baselines including DeepSynergy (71.6\% AUROC, 78.8\% AUPRC), BAITSAO (69.5\% AUROC, 75.6\% AUPRC), and HERMES (50.0\% AUROC). This performance advantage in high-order combinatorial spaces highlights the framework's ability to capture complex, context-dependent drug interactions through its heterogeneous graph representation.

\subsection{Limitations and Future Directions}
Despite its strong performance, SynCell has several limitations that warrant consideration. First, while we incorporate protein-protein interactions, drug targets, and gene expression data, additional biological information—including metabolic pathways, signaling networks, and detailed drug structural features—could further enhance predictive accuracy. Future work should explore integration of multi-omics data and comprehensive chemical representations to capture more complete biological contexts \cite{wang2024clinical}. 

Second, our model relies on cell line-derived expression profiles from DepMap, which may not fully capture the tumor microenvironment and heterogeneity present in patient tissues. While cell lines provide a controlled system for mechanistic discovery, discrepancies in stromal interactions and immune contexture remain a challenge for clinical translation \cite{zhang2022shifting, arafeh2025depmap}. Future iterations could integrate patient-derived xenograft (PDX) data or single-cell resolved spatial transcriptomics to bridge this gap.

Third, our framework operates on static biological networks with fixed expression thresholds (e.g., 80th percentile), whereas cellular responses involve dynamic processes. Incorporating temporal dynamics of pathway activation and gene expression changes would provide more mechanistic insights into synergy mechanisms and enable prediction of time-dependent drug effects, a direction recently highlighted in foundation models for precision medicine \cite{huang2024txgnn}. Finally, while the heterogeneous graph structure provides inherent interpretability through explicit biological relationships, developing specialized interpretation mechanisms to identify key proteins and pathways responsible for specific synergy predictions would enhance practical utility for guiding experimental validation and clinical translation.

\subsection{Broader Implications}
SynCell's capability to accurately predict synergy for novel drug combinations has significant implications for drug discovery and repurposing. The exceptional performance in DrugSingle and DrugDouble scenarios suggests potential applications in early-stage drug development where limited experimental data is available for new chemical entities. This could substantially reduce the cost and time required for experimental screening of combination therapies, accelerating therapeutic development amidst the growing global cancer burden \cite{bray2024global}. 

Our work complements recent advances in zero-shot drug repurposing, such as TxGNN \cite{huang2024txgnn}. While TxGNN focuses on identifying \textit{single-agent} indications for diseases with limited treatments, SynCell addresses the \textit{combinatorial} space, predicting synergistic \textit{pairs} tailored to specific molecular contexts. Together, these approaches form a comprehensive toolkit: TxGNN expands the pool of potential therapeutic candidates, and SynCell optimizes their combinatorial usage to overcome resistance and reduce toxicity.

The cross-dataset generalization demonstrated on DrugComb further validates the framework's practical utility, indicating that the biological knowledge encoded in the graph structure transfers effectively across different experimental settings and measurement protocols \cite{liu2020drugcombdb}. This robustness is crucial for real-world applications where model performance must be maintained across diverse data sources and experimental conditions. Furthermore, the successful integration of heterogeneous biological data within a unified graph framework establishes a flexible foundation that can be readily extended to incorporate additional data types, such as single-cell expression profiles, proteomics data, or clinical patient information. This adaptability positions SynCell for applications in personalized medicine, where patient-specific genomic and molecular profiles could be incorporated to predict individualized drug combination responses. As combination therapies become increasingly important for treating complex diseases, computational frameworks that can reliably predict synergy for novel compounds will play a crucial role in accelerating therapeutic development and optimizing treatment strategies. SynCell represents a significant step toward this goal, offering both predictive accuracy and mechanistic interpretability for guiding precision oncology approaches.

\bibliographystyle{unsrt}
\bibliography{main}

\end{document}